# Automatic Classification of Music Genre using Masked Conditional Neural Networks

Fady Medhat     David Chesmore     John Robinson
Department of Electronic Engineering
University of York
York, United Kingdom
{fady.medhat, david.chesmore, john.robinson}@york.ac.uk

*Abstract*—Neural network based architectures used for sound recognition are usually adapted from other application domains such as image recognition, which may not harness the time-frequency representation of a signal. The ConditionaL Neural Networks (CLNN) and its extension the Masked ConditionaL Neural Networks (MCLNN)[1] are designed for multidimensional temporal signal recognition. The CLNN is trained over a window of frames to preserve the inter-frame relation, and the MCLNN enforces a systematic sparseness over the network's links that mimics a filterbank-like behavior. The masking operation induces the network to learn in frequency bands, which decreases the network susceptibility to frequency-shifts in time-frequency representations. Additionally, the mask allows an exploration of a range of feature combinations concurrently analogous to the manual handcrafting of the optimum collection of features for a recognition task. MCLNN have achieved competitive performance on the Ballroom music dataset compared to several hand-crafted attempts and outperformed models based on state-of-the-art Convolutional Neural Networks.

## I. INTRODUCTION

Statistical methods have been used in sound recognition for decades. Especially, the GMM-HMM combination, where the GMM is used to model the statistical distribution of the frames and the HMM to capture the state transitions in the sequential signal. These models have been used extensively for speech [1]. Despite the success of such models they involve a time-consuming stage of manual inspection of the space of features that can be extracted from the raw signal with an objective of finding the most distinctive features to introduce to a recognition model.

Deep neural network architectures are gaining momentum in an endeavor to eliminate the need to handcraft the features required for classification. A remarkable attempt for image recognition was in [2] using a deep Convolutional Neural Network (CNN) [3]. In parallel, efforts were considered to apply these architectures for sound recognition [4],[5],[6],[7].

Widely adapted models such as Convolutional Neural Networks and Deep Belief Nets (DBN) [8] were not initially designed for sound, but rather they got adapted to sound recognition after their success in image processing, which may not harness the nature of the sound signal in a time-frequency representation such as a spectrogram. For example, the DBN does not consider the inter-frame relation of a temporal signal, and the CNN depends on weight-sharing, which does not preserve the spatial locality of the learned features across the frequency bins in a spectrogram.

The ConditionaL Neural Networks (CLNN) [9] take into consideration the inter-frame relation across a temporal signal. The Masked ConditionaL Neural Networks (MCLNN) [9] extend upon the CLNN structure by embedding a filterbank-like behavior within the network, which encourages the network to learn in frequency bands and supports sustaining frequency-shifts. Additionally, the mask is designed to allow a concurrent exploration of a range of feature combinations analogous to the exhaustive manual search for finding the optimum combination of features. In this work, we explore using a shallow architecture of the MCLNN with a wider segment compared to the deep architecture used in [10] for music genre classification.

The rest of the paper is organized as follows: In the next section, Section II, we will refer to other related models. Section III will discuss the ConditionaL Neural Network (CLNN) and a follow on is brought in Section IV to discuss the Masked Conditional Neural Network that extends upon the CLNN. Section V will go through the experiments with their relevant discussion of the findings. Finally, we wrap up this work in Section VI with the conclusions and future work.

## II. RELATED MODELS

The Restricted Boltzmann Machine (RBM) [11] is an architecture of two layers of neurons, a visible and a hidden layer, with bi-directional connections going across the two layers. An RBM is trained generatively, aiming to model the feature space of the training data. Several RBMs can be stacked on top of each other to form a Deep Belief Net structure (DBN) [8]. Each layer of an RBM extracts a more abstract representation than the layer before it, which is a property for deep architectures in general. This enhances the features extracted and consequently the classification decision. Three RBMs were stacked by Hamel et al. [7] to extract the features from audio signals of music genres that were further classified

---

[1] Code: https://github.com/fadymedhat/MCLNN

This work is funded by the European Union's Seventh Framework Programme for research, technological development and demonstration under grant agreement no. 608014 (CAPACITIE).

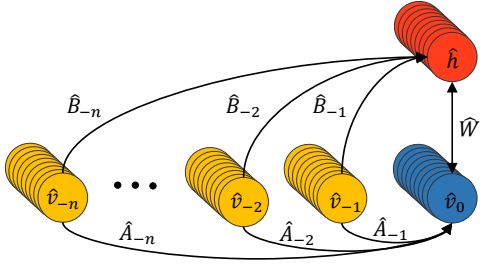

Fig. 1. The Conditional RBM structure

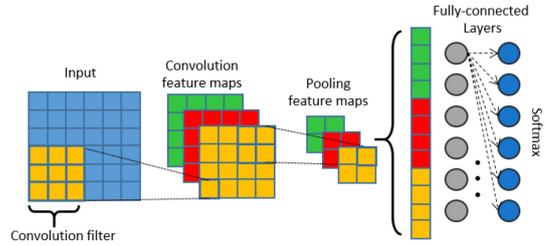

Fig. 2. Convolutional Neural Network

using an SVM. In their work, they showed the abstraction captured by each layer of an RBM, where using the three layers achieved higher accuracy compared to using a double or a single layer of an RBM when training the model on frame level of a spectrogram. Despite the interesting findings, RBMs ignore the temporal aspect of a signal by treating each frame in isolation.

The Conditional Restricted Boltzmann Machine (CRBM) [12] by Taylor et al. was introduced to allow the RBM to be adapted for temporal signals. The CRBM involves the inclusion of conditional links from the previous input states to the current hidden one and auto-regressive links to the current input. Accordingly, the prediction of the hidden layer's activations and the reconstructed vector at the visible layer are conditioned on the past $n$ frames. Fig. 1 shows a CRBM structure. The vanilla RBM is represented with its visible $\hat{v}_0$ and hidden $\hat{h}$ nodes with the bidirectional connections $\hat{W}$ going across the visible and hidden layers. The CRBM's conditional links are represented by the directed connections from the previous visible states $\hat{v}_{-1}$, $\hat{v}_{-2}$, …, $\hat{v}_{-n}$ to the current hidden layer $\hat{h}$ using $\hat{B}_{-1}$, $\hat{B}_{-2}$, …, $\hat{B}_{-n}$ and to the current visible state $\hat{v}_0$ using $\hat{A}_{-1}$, $\hat{A}_{-2}$, …, $\hat{A}_{-n}$. The CRBM was applied on a multi-channel temporal signal for modelling the human motion through joints tracking. Later attempts adapted the CRBM to the phoneme recognition task by Mohamed et al. [13] and for drum pattern analysis by Battenberg et al. in [14]. They were also extended in the Interpolating CRBM (ICRBM) [13] by including the future frames' influence in addition to the past ones. ICRBM outperformed the CRBM for the phoneme modelling in speech recognition.

Fully connected Feed-forward Neural Networks do not scale well with inputs of large dimensions such as images. Convolutional Neural Networks (CNN) [3] use weight sharing to tackle this problem. The CNN architecture as shown in Fig. 2 is based on the hypothesis that the features detected in one region of a 2-dimensional input have a high probability of being detected elsewhere in the image. Therefore, using small filters (weight matrices of small sizes, e.g. 5×5) to scan the image, a CNN does not require a direct connection between each pixel in the input and the hidden layer. The filters scan the image, where each filter acting as an edge detector extracts different properties with respect to its neighboring filters and project it on a new feature map representation (a new version of the original image). The generated feature maps are rescaled through a pooling operation, where a small window extracts the mean or the max value across the pixels in the window to project it over a lower resolution feature map. These two operations of convolution and pooling are interleaved several times to form deep architectural structures, where the final feature maps are flattened to a single feature vector to be fed to a fully connected network as depicted in Fig. 2. CNN achieved remarkable results for image recognition in [2].

The spatial locality of the energy captured at different frequency bins is a distinctive property to the sound category. The weight sharing allows the CNN to be translation invariant, which does not take into consideration the nature of the time-frequency representation. Accordingly, several attempts [5],[4],[6],[15] tried to tailor the CNN to fit the nature of the multi-dimensional temporal representation of a sound signal. For example, the work by Abdel-Hamid et al. [5] redesigned the weight sharing to a *Limited Weight Sharing*, which limits the sharing to a particular region of frequencies. Other attempts such as the work by Pons et al. [15] combined two sets of filters in one model to scan the temporal and the spectral dimensions separately, which achieved a higher accuracy compared to a normal CNN.

III. CONDITIONAL NEURAL NETWORKS

The ConditionaL Neural Network (CLNN) [9] is a discriminative model extending from the generative CRBM discussed earlier. The CLNN adopts one set of the CRBM's conditional links, which connect the previous visible states to the hidden layer. The CLNN also extends the influence of the frames to the future in addition to the past frames as in the ICRBM. The CLNN is trained over a window of frames, where it predicts the window's middle frame conditioned on $n$ frames on both temporal directions.

The CLNN has a hidden layer of $e$ neurons of a vector shape. The input to the CLNN is a window having $d$ frames. The width of the window follows (1)

$$d = 2n + 1 \qquad (1)$$

where the width $d$ is specified by twice the order $n$ (to account for future and past frames) in addition to the window's middle frame. The order $n$ specifies the number of frames in a single temporal direction. There are dense connections between each frame in the window and hidden layer, where the activation of single neuron is formulated in (2)

$$y_{j,t} = f\left(b_j + \sum_{u=-n}^{n}\sum_{i=1}^{l} x_{i,u+t} W_{i,j,u}\right) \qquad (2)$$

where the $j^{th}$ hidden node activation $y_{j,t}$ (the index $t$ specifies the position of the frame in the input segment discussed later) is given by the output of the transfer function $f$. The bias at the node is $b_j$. The $x_{i,u+t}$ is the $i^{th}$ feature of feature vector $x$ of length $l$ at position $u + t$ in a window, where the $u$ ranges between [-$n$, $n$] and the frame at $u = 0$ is the window's middle frame at position $t$ of a segment. $W_{i,j,u}$ is the weight between the $i^{th}$ feature and the $j^{th}$ hidden node of the matrix at position $u$ in the weight tensor. The window of frames is extracted from a larger temporal chunk that we will refer to as the segment. The segment has a minimum width of a window and can be larger as we will discuss later. The activation can be reformulated in a vector form in (3)

$$\hat{y}_t = f\left(\hat{b} + \sum_{u=-n}^{n} \hat{x}_{u+t} \cdot \widehat{W}_u\right) \quad (3)$$

where $\hat{y}_t$ is the activation pattern at the hidden layer. $f$ is the transfer function. $\hat{b}$ is the bias vector. $\hat{x}_{u+t}$ is the vector at index $u + t$, where $u$ is within the interval [-$n$, $n$]. $\widehat{W}_u$ is the matrix at index $u$ of the weight tensor of dimensions [feature vector length $l$, the hidden layer width $e$, the number of frames in a window $d$]. Accordingly, for each frame in the window a dedicated matrix of the tensor is used to process it. The output of the vector-matrix multiplication is $d$ vectors each of length $e$ to be summed together feature-wise before the nonlinearity applied by the transfer function. The conditional distribution of the window's middle frame conditioned on the $n$ frames on both sides is captured by $p(\hat{y}_t | \hat{x}_{-n+t}, ... \hat{x}_{-1+t}, \hat{x}_t, \hat{x}_{1+t}, ... \hat{x}_{n+t}) = \sigma(...)$, where $\sigma$ is a Sigmoid or the output layer's Softmax.

Fig. 3 shows two CLNN layers scanning a multi-dimensional temporal signal. The output frames at each CLNN layer are decremented by $2n$ frames. To account for this reduction in the number of frames, the input to a deep CLNN architecture is a chunk of frames referred to as the segment. The segment size follows (4)

$$q = (2n)m + k, \text{ where } n, m \text{ and } k \geq 1 \quad (4)$$

where a segment $q$ depends on the order $n$ (the 2 is for the future and past frames), $m$ is the number of layers and $k$ is for the extra frames that should remain after the CLNN layers. These $k$ frames can be flattened to a single vector or pooled across (e.g. mean or max pooling) as in a CNN, but for sound, it is a single dimension pooling. For example, at $n = 4$, $m = 3$ and $k = 5$, the input segment at the first layer (of the three layers at $m=3$) is of size (2×4)×3+5 = 29. The output at the first layer is 29 – (2×4) = 21 frames. The output of the second layer is 21 – (2×4) = 13 frames and finally the output of the third layer is 13 – (2×4) = 5 frames. The remaining 5 frames represent the extra frames to be flattened or pooled across before transferring them to the fully connected layers for the final classification.

IV. MASKED CONDITIONAL NEURAL NETWORK

Spectrograms provide an insight of the energy contribution at each frequency bin as the sound signal progresses through time. The energy of a single frequency bin may smear across nearby bins due to uncontrolled propagation factors, which affect the spectrogram representation. This alteration is not helpful for recognition systems. A filterbank is a group of filters used to subdivide the spectrogram representation into frequency bands. The partitioning provides a representation that counters the smearing, which allows the spectrogram to be frequency shift-invariant by aggregating the energy across several frequency bins. Additionally, the transformed spectrogram has fewer dimensions matching the number of filters. A filterbank is the main operating block in scaled spectrograms using e.g. Bark or Mel scale. For example, in the Mel-scaled filterbank, the filters are scaled by having their center frequencies Mel-spaced from each other.

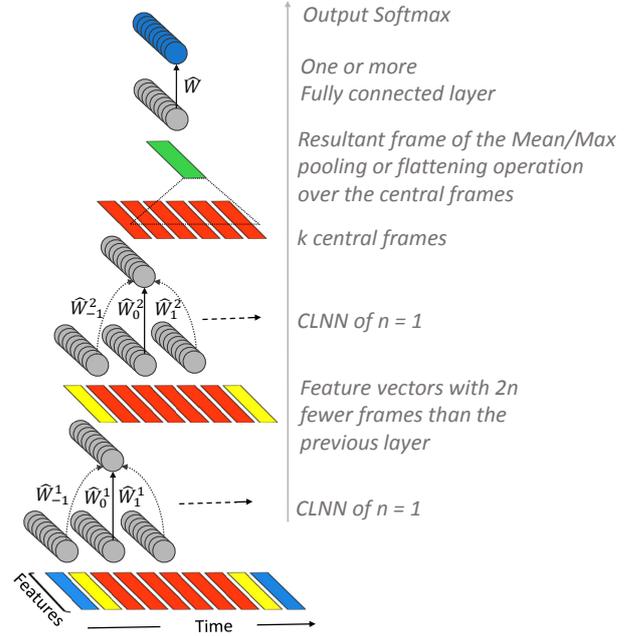

Fig. 3 A two layer CLNN model with $n=1$

The Masked ConditionaL Neural Network (MCLNN) [9] embeds a filterbank-like behavior within the network by enforcing a systematic sparseness over the network's links. The band-like sparseness induces the network to learn in frequency bands and permit each hidden node to be an expert in a localized region of the feature vector, which allows focusing on distinctive features in the node's field of observation.

The mask is a binary pattern as depicted in Fig. 4. It is designed to follow a band-like structure using two tunable hyper-parameters: the Bandwidth and the Overlap. The Bandwidth specifies the number of column-wise 1's, which refers to the number of features that will be considered together from a feature vector. The Overlap controls the superposition distance between two successive columns. For example, Fig. 4.a depicts a mask having a *Bandwidth=5* and an *Overlap=3*. The positions of the ones activate different regions within the feature vector. Fig. 4.b shows the enabled links with respect to the hidden layer nodes matching the pattern enforced in Fig. 4.a. Fig. 4.c shows an example of a masking of a *Bandwidth = 3* and an *Overlap = –1*, where a negative Overlap refers to the non-overlapping distance between two consecutive columns. The linear spacing of the 1's patterns is specified using

$$lx = a + (g - 1)(l + (bw - ov)) \quad (5)$$

where the linear index $lx$ is specified by the length of the vector $l$, the bandwidth $bw$ and the overlap $ov$. The values of $a$ ranges

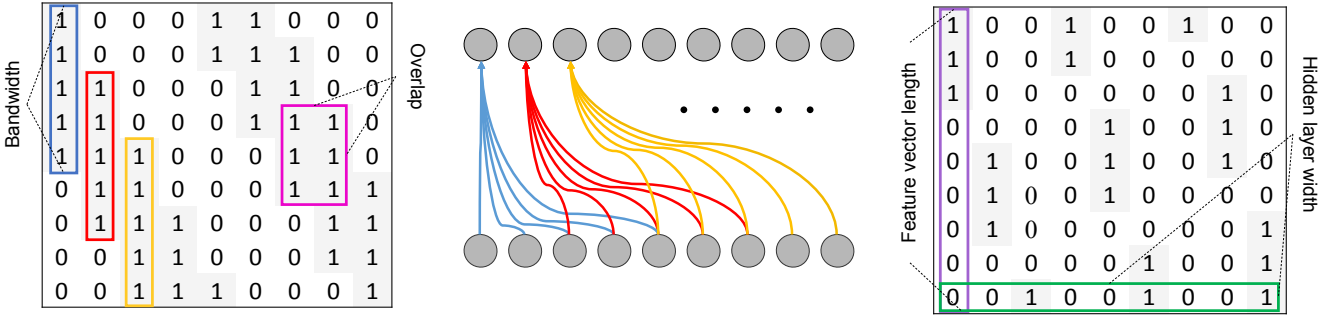

Fig. 4 Examples of the Mask patterns. a) A bandwidth of 5 with an overlap of 3, b) The allowed connections matching the mask in a. across the neurons of two layers, c) A bandwidth of 3 and an overlap of -1

between [0, $bw$-1] and $g$ is in the range $[1, \lceil (l \times e)/(l + (bw - ov)) \rceil ]$, where $e$ is the hidden layer width.

In addition to the mask's role in imposing a band-like structure over the links, it automates the exploration of a range of feature combination concurrently analogous to mixing-and-matching the optimum combination of features manually for a recognition task. Fig. 4.c shows three shifted versions of a filterbank-like pattern depicted in the 1st three columns compared to the 2nd batch of three columns and the 3rd ones. Each filterbank-like version is aggregating different sets of feature combinations, and all the aggregated sets are considered concurrently. On a granular level, the input at the 1st hidden node (mapped to the first column) in Fig. 4.c is the first three features disregarding the temporal dimension $t$ at this stage. The input at the 4th node is the first two features and at the 7th node's input is the first feature only.

The masking operation is imposed over the network's connection through an element-wise multiplication between mask's matrix and each weight matrix in the tensor following (4)

$$\hat{Z}_u = \widehat{W}_u \circ \widehat{M} \quad (6)$$

where $\widehat{W}_u$ is the original weight matrix at index $u$, $\widehat{M}$ is the masking pattern and $\hat{Z}_u$ is the masked version of the weight matrix to substitute $\widehat{W}_u$ in (3).

Fig. 5 shows a single step of the MCLNN scanning a window of frames having order $n$. The window size is $2n+1$. Accordingly, the weight tensor has a depth of $2n+1$, where each frame in the input window is processed with its corresponding weight matrix at the same index. The highlighted regions in the mask are the active connections following the locations of the 1's. The resultant of the depicted MCLNN step is a single vector representation for the window of frames.

## V. EXPERIMENTS

The experiments use a shallow architecture of the MCLNN compared to the deep model used in [10]. We evaluated the MCLNN performance on the music genre classification using the Ballroom [16] dataset. The dataset is composed of 698 music clip of 30 seconds each for 8 Ballroom music sub-genres: Cha Cha, Jive, Quickstep, Rumba, Samba, Tango, Viennese Waltz and Slow Waltz. As an initial processing step, we resampled the files to a monaural 16-bit word depth wav format at a sampling rate of 22050 Hz. All files underwent a spectrogram transformation to a 256 frequency bin logarithmic mel-scaled spectrogram with an FFT window of 2048 sample and 50% overlap. Further preprocessing involved extracting segments following (4). We followed a 10-fold cross-validation to report the accuracies, where the training folds are standardized, and the z-score parameters are applied on the validation and test folds.

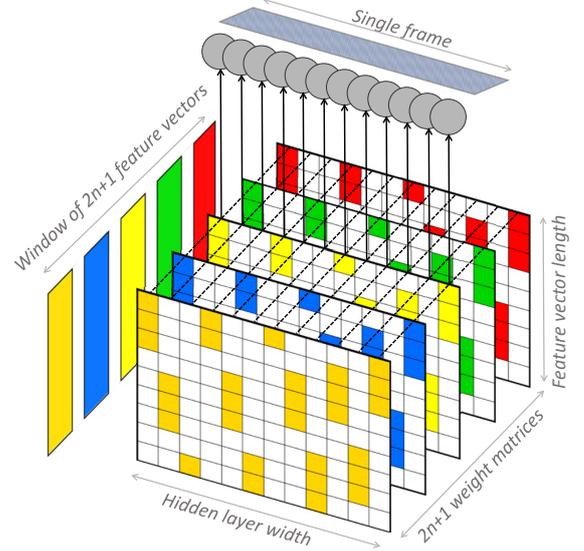

Fig. 5 A single step of the MCLNN

TABLE I. MCLNN HYPER-PARAMETERS FOR THE BALLROOM

| Layer | Type | Nodes | Mask Bandwidth | Mask Overlap | Order $n$ |
|---|---|---|---|---|---|
| 1 | MCLNN | 220 | 40 | -10 | 20 |

We adopted a single layer MCLNN followed by a pooling layer to pool across $k$=55 frames in addition to the window's middle frame after the MCLNN layers and two fully connected layers of 50 and 10 neurons, respectively, before the final 8-way output softmax. The total segment length at an order $n = 20$ is 96 frames. The model hype-rparameters are listed in Table I.

The model is trained to minimize categorical cross-entropy between the prediction of a segment and the labels using ADAM [17]. Dropout [18] was used for regularization. The category of a music file was decided using probability voting across the predictions of the frames of a clip.

TABLE II. PERFORMANCE ON BALLROOM DATASET USING MCLNN COMPARED WITH ATTEMPTS IN THE LITERATURE

| Classifier and Features | Acc. % |
|---|---|
| SVM + 28 feature *with* Tempo [19] | 96.13 |
| KNN + Modulation Scale Spectrum [20] | 93.12 |
| Manhattan Distance + Block-Level features [21] | 92.44 |
| **MCLNN (Shallow, *n*=20, *k*=55) + Mel-Spec. (this work)** | **92.12** |
| MCLNN (Deep, *n*=15, *k*=10)+ Mel-Spec. [10] | 90.40 |
| SVM + Rhyth., Hist., Statist., Onset, Symb. [23] | 90.40 |
| KNN + 15 MFCC-like descriptors *with* Tempo [22] | 90.10 |
| KNN + Rhythm and Timbre [24] | 89.20 |
| SVM + 28 features *without* Tempo [19] | 88.00 |
| CNN+ Mel-Scaled Spectrogram[15] | 87.68 |
| SVM + Rhyth. + Hist. + Statist. features [25] | 84.20 |
| KNN + Tempo [22] | 82.30 |

Table II lists the accuracies reported on the Ballroom dataset through several attempts in the literature including the MCLNN. The highest accuracy in the table is reported in the work of Peeters [19], where he achieved 96.13% using his proposed handcrafted features with the help of the Tempo annotations released with the dataset. Peeters reapplied his method without the tempo annotations, and the accuracy dropped to 88%. An accuracy of 93.12% and 92.44% was also reported on the Ballroom dataset in [20] and [21], respectively, where both attempts exploited handcrafted features. The effectiveness of the tempo annotations was also studied by Gouyon et al. [22], where they achieved a base accuracy of 82.3% using the tempo data only. They achieved an accuracy of 90.1% using the tempo annotations in combination with their proposed handcrafted features. Conversely, the MCLNN achieved 92.12% without any handcrafted features. Additionally, there is no special tailoring to exploit musical perceptual properties or tempo annotations, which allows the MCLNN to be considered for other multi-channel temporal representations.

In comparison to a neural based architecture, the work of Pons et al. [15] achieved 87.68% using a specially designed shallow Convolutional Neural Network architecture of musically motivated filters. Their architecture used a single convolution layer followed by a max polling layer before a 200 neuron fully connected layer and the 8-way Softmax. In their work, they compared the performance of three CNN architectures based on the size of the filter used: Black-box, Frequency filters and Temporal filters. The Black-box filters are of size $m \times n$, which are the regular CNN filters, the Frequency filters are of size $m \times 1$ to convolve the frequency dimension and $1 \times n$ sized filters for the temporal dimension. Table III lists the accuracies of 10-fold cross validation achieved in the work of Pons et al. in addition to the accuracy achieved by the MCLNN. The highest accuracy achieved by Pons et al. was obtained when combining the filters trained separately across each of temporal and frequency dimensions in the same model. Their combined architecture achieved 86.54%, which increased to 87.68% when using a pre-trained model. As the table shows, the MCLNN achieved the highest accuracy with the lowest standard deviation, which demonstrates the stability of the accuracy reported by the MCLNN across the folds.

TABLE III. MCLNN COMPARED WITH PONS ET AL. [15] CONVOLUTIONAL NEURAL NETWORK PERFORMANCE FOR THE BALLROOM

| Classifier and Features | Acc. % ± Std. |
|---|---|
| **MCLNN + Mel-Scaled Spectrogram (this work)** | **92.12 ± 2.94** |
| Time-Frequency pre-trained CNN | 87.68 ± 4.44 |
| Black-Box CNN | 87.25 ± 3.39 |
| Time-Frequency CNN | 86.54 ± 4.29 |
| Time Filter - CNN | 81.79 ± 4.72 |
| Frequency Filter - CNN | 59.59 ± 5.82 |

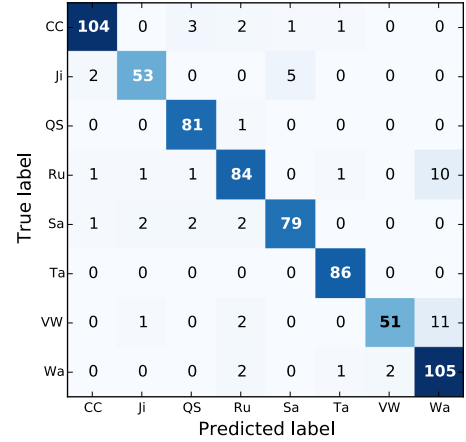

Fig. 6. Ballroom confusion using MCLNN. Cha Cha (CC), Jive (Ji), Quickstep (QS), Rumba (Ru), Samba (Sa), Tango (Ta), Viennese Waltz (VW) and Slow Waltz (Wa)

Fig. 6 shows the confusion across the Ballroom genres using the MCLNN. There is high confusion rate between the Rumba and the Waltz similarly for the Viennese Waltz and the Waltz. Lykartsis et al. reported similar confusion trends in [26].

The performance of the MLCNN is accounted to exploiting both dimensional representations, either the temporal or the frequency dimension. Additionally, the masking operation counter the smearing that occurs naturally across frequency bins due to uncontrolled circumstances related to the sound propagation. Through various Overlap and Bandwidth settings, we noticed that increasing the sparseness through negative Overlap values, enhanced the accuracy. This is accounted to disabling the effect of the smearing early on before the smearing noise across the frequencies propagates into the network and affect the classification decision. MCLNN accuracy has surpassed several handcrafted attempts in addition to state-of-the art Convolutional Neural Networks. Additionally, the MCLNN have achieved competitive results compared to handcrafted features without using any musical perceptual properties exploited by other efforts for music tasks. This provides a promising argument to the generalization of the MCLNN to multidimensional temporal signals other than spectrograms, which we will consider in future work.

VI. CONCLUSIONS AND FUTURE WORK

We have explored the ConditionaL Neural Network (CLNN) and the Masked ConditionaL Neural Network designed for multidimensional temporal signal recognition. The CLNN preserves the inter-frame relations, and the MCLNN extends upon the CLNN by enforcing a masking operation over the network's links that follows a band-like representation mimicking a filterbank. The masking process induces the network to learn in frequency bands and assist a neuron in the hidden layer to be an expert in a localized region of the feature vector. The mask also automates the exploration of a range of feature combinations concurrently, analogous to the manual optimization of mixing and matching different feature combinations for a recognition task. MCLNN outperformed several state-of-the-art handcrafted and Convolutional Neural Networks based attempts for the music genre recognition task using the Ballroom music genre dataset. MCLNN achieved these accuracies without applying any augmentations or depending on any musical perceptual features used by other attempts. Future work will explore different MCLNN architectures with optimized masking patterns and different order $n$ across the MCLNN layers. Future work will consider applying the MCLNN to other multi-channel temporal signal representations other than sound.